\documentclass{emulateapj}
 \usepackage{epstopdf}

\shorttitle{Identification of a Wide, Low-Mass Multiple System}
\shortauthors{Faherty et al.}

\begin{document}


\title{Identification of a Wide, Low-Mass Multiple System Containing the Brown Dwarf 2MASS J0850359+105716\footnote{T\lowercase{his paper includes data gathered with the 6.5m Magellan Telescopes
located at Las Campanas Observatory, Chile.}}}


\author{Jacqueline K.\ Faherty\altaffilmark{1}} 
\affil{Department of Astrophysics, American Museum of Natural History, Central Park West at 79th Street, New York, NY 10034; jfaherty@amnh.org }

\author{Adam J.\ Burgasser\altaffilmark{2} }
\affil{Center for Astrophysics and Space Science, University of California San Diego, La Jolla, CA 92093, USA}

\author{John J.\  Bochanski}
\affil{Department of Astronomy and Astrophysics, The Pennsylvania State University, University Park, PA 16802, USA}

\author{Dagny L.\ Looper\altaffilmark{3} }
\affil{Institute for Astronomy, University of Hawai'i, 2680 Woodlawn Dr, Honolulu, HI 96822, USA}

\author{Andrew A.\  West}
\affil{Department of Astronomy, Boston University, 725 Commonwealth Ave Boston, MA 02215.}

\author{Nicole S.\ van der Bliek}
\affil{CTIO/National Optical Astronomy Observatory, Casilla 603, La Serena, Chile}

\altaffiltext{1}{Visiting astronomer, Cerro Tololo Inter-American Observatory, National Optical Astronomy Observatory, which are operated by the Association of Universities for Research in Astronomy, under contract with the National Science Foundation.}
\altaffiltext{2}{Hellman Fellow.}
\altaffiltext{3}{Visiting astronomer, IRTF.}

\begin{abstract}
We report our discovery of NLTT 20346 as an M5+M6 companion system to the tight binary (or triple) L dwarf 2MASS J0850359+105716.  This nearby ($\sim$31 pc), widely separated ($\sim$7700 AU) quadruple system was identified through a cross-match of proper motion catalogs.  Follow-up imaging and spectroscopy of NLTT 20346 revealed it to be a magnetically active M5+M6 binary with components separated by $\sim$2$\arcsec$ (50-80 AU).   Optical spectroscopy of the components show only moderate H$\alpha$ emission corresponding to a statistical age of $\sim$5 - 7 Gyr for both M dwarfs. However NLTT 20346 is associated with the  XMM-Newton source J085018.9+105644, and  based on X-ray activity the age of NLTT 20346 is between 250-450 Myr. Strong Li absorption in the optical spectrum of 2MASS J0850+1057 indicates an upper age limit of 0.8 - 1.5 Gyr favoring the younger age for the primary.    Using evolutionary models in combination with an adopted system age of 0.25-1.5 Gyr indicates a total mass for 2MASS J0850+1057 of 0.07$\pm$0.02 M$_{\sun}$ if it is a binary.  NLTT 20346/2MASS J0850+1057 joins a growing list of hierarchical systems containing brown dwarf binaries and is among the lowest binding energy associations found in the field. Formation simulations via gravitational fragmentation of massive extended disks have successfully produced a specific analog to this system.

\end{abstract}

\keywords{Astrometry-- stars: low-mass-- brown dwarfs-- stars: fundamental parameters--binaries: wide-- stars: individual  (2MASS J08503593+1057156, NLTT 20346)}
\section{INTRODUCTION}

Very low--mass stars and brown dwarfs (VLMs and BDs; M $\lesssim$ 0.10 M$_{\sun}$) are among the most populous constituents of the Galaxy yet their origins remain controversial.  Whether they form in a manner similar to higher mass stars or require additional or completely different processes is currently under debate (see \citealt{2007prpl.conf..427B}; \citealt{2007prpl.conf..459W}; \citealt{2007prpl.conf..443L} and references there-in).  One important characteristic that indicates a difference in formation is their binary frequency and multiplicity statistics (separation, mass ratio, eccentricities, etc).  Contrary to higher mass stars, VLM binaries are found tightly bound ($\rho$ $<$ 20 AU), in small frequency  (10-20$\%$) and with high mass ratios (typically q$\sim$1 -- e.g. \citealt{2003AJ....126.1526B}, \citealt{2003ApJ...587..407C}, \citealt{2003ApJ...586..512B}, \citealt{2007ApJ...671.2074A}, \citealt{2008AJ....135..580R}).  

Over the past decade, theoretical and observational studies have converged on several competing mechanisms to explain the formation of VLM stars and brown dwarfs.   Among the most prominent are (1) magnetoturbulent or gravoturbulent fragmentation (\citealt{2004ApJ...617..559P}; \citealt{2003MNRAS.339..577B}; \citealt{2004A&A...414..633G,2004A&A...423..169G,2006A&A...452..487G}); (2) ejection of a protostellar embryo from the natal core (\citealt{2001AJ....122..432R}; \citealt{2005MNRAS.356.1201B}); or (3) disk fragmentation (\citealt{2006A&A...458..817W}; \citealt{2009arXiv0911.3662S}).  Characteristics of multiple systems predicted by each have been used to differentiate which mechanism is most probable for the population.  Early versions of the ejection model were favored in part because they predicted that brown dwarf binaries  should have separations no greater than 10 AU (\citealt{2002MNRAS.332L..65B}). This was supported through early observational studies. However recent findings of widely separated VLM systems (typically $\rho$ $>$ 100 AU) are contradictory and shed doubt on the viability of the ejection scenario (\citealt{2004ApJ...614..398L}; \citealt{2009ApJ...697..824A}; \citealt{2007ApJ...660.1492C}; \citealt{2009ApJ...691.1265L}; \citealt{2005A&A...440L..55B}; \citealt{2007ApJ...667..520C}; \citealt{2007ApJ...659L..49A}; \citealt{2009arXiv0903.3251R}).  Updated simulations have successfully created a handful of wide brown dwarf binaries (\citealt{2005MNRAS.356.1201B,2009MNRAS.392..590B}) but the growing number in the field indicates a fraction too high to be explained by just this mechanism.  The existence and frequency of wide substellar pairs may not discount one mechanism versus another; rather multiple formation mechanisms may be at work in the field.


Recent work has shown that brown dwarfs widely separated from nearby stars have a higher frequency of also being tight multiples (\citealt{2005AJ....129.2849B}; \citealt{2009AJ....137....1F}).  This has been explored in the disk fragmentation models of \citet{2009arXiv0911.3662S} where similar high frequencies of brown dwarf multiples were produced when widely separated from a more massive companion.   In this article, we report the identification of another widely separated brown dwarf binary (or triple; see Burgasser et al 2010) companion, the NLTT 20346 and 2MASS J0850+1057 comoving system.   In Section 2, we review the discovery and observational data used to characterize the system.  In Section 3, we discuss the astrometric details of the system including a contaminating artifact that likely skewed previous parallax measurements for 2MASS J0850+1057.  In Section 4, we analyze the details of the system including the activity and masses of  each component. In Section 5, we discuss its importance among the population of VLM wide companion systems.  Results are summarized in Section 6.

\section{OBSERVATIONS}
{\it \subsection{Targets}}
2MASS J0850359+105716 (hereafter 2MASS J0850+1057) was first identified as a single object in the Two Micron All Sky Survey (2MASS; \citealt{2006AJ....131.1163S}) by \citet{1999ApJ...519..802K}.  It was subsequently assigned to be the prototype for the L6 spectral subclass before being resolved as an 0$\farcs$16 binary system by \citet{2001AJ....121..489R}. \citet{2008A&A...481..757B} confirmed the binarity of 2MASS J0850+1057 through common proper motion detected in multiple epoch Hubble Space Telescope (HST) data.   The magnitude difference between components of 2MASS J0850+1057 in HST images indicates that the secondary is a late-type L dwarf (\citealt{2001AJ....121..489R}; \citealt{2008A&A...481..757B}).   Burgasser et al. 2010 report template fitting to the combined light spectrum of 2MASS J0850+1057 and conclude component spectral types of L7 and L6.5.  They also speculate that the brighter but later-type primary may itself be a closely-separated, unresolved pair, making this system one of a handful of candidate brown dwarf triples (e.g. Gliese 569, \citealt{2006ApJ...644.1183S}; DENIS J0205-1159, \citealt{2005AJ....129..511B}; Kelu 1, \citealt{2009AIPC.1094..561S}).

NLTT 20346 was  identified as a high proper motion star in \citet{1979lccs.book.....L} and subsequently in \citet{2002AJ....124.1190L}.   \citet{1979lccs.book.....L}  assigned it a spectral class of K using color information, but little else was known about NLTT 20346 prior to this work.  

{\it \subsection{ ISPI Imaging}}
We observed the field containing 2MASS J0850+1057 and NLTT 20346 at the Cerro Tololo 4.0m Blanco telescope six times between 2008 March 09 and 2010 April 08 as part of an ongoing astrometric program.  The Infrared Side Port Imager (ISPI) was used and the target was observed through the $J$ band filter (\citealt{2004SPIE.5492.1582V}).  ISPI has a nominal plate scale of 0.3$\arcsec$/pixel. Five images of 60 s exposures with 4 co-adds were obtained at each of the five dither positions.  Observations were made under clear conditions with seeing ranging from $\sim$0.7$ - $1.0$\arcsec$.

Dark frames and lights on/off dome flats were obtained at the start of each evening.  The subsequent reduction procedures were based on the prescriptions put together by the ISPI team\footnote{$http://www.ctio.noao.edu/instruments/ir\_instruments/ispi/$} utilizing a combination of IRAF routines as well as publicly available software packages: e.g. WCSTOOLS (\citealt{2002ASPC..281..169M}).  $J$ band flats were created by median combining the lights on and lights off images then subtracting the two.  Bad pixel masks were created from a dome flat lights on and lights off image.  Individual frames were flat-fielded and corrected for bad pixels with the resultant calibration images.  All images were flipped to orient North up and East to the left using the IRAF routine $\it{osiris}$ in the $\it{cirred}$ package.  Finally, the IRAF routine $\it{xdimsum}$ was used to perform sky subtractions and mask resultant holes from bright stars.  

\begin{figure}[htbp]
\centering
\epsscale{1.0}
\plotone{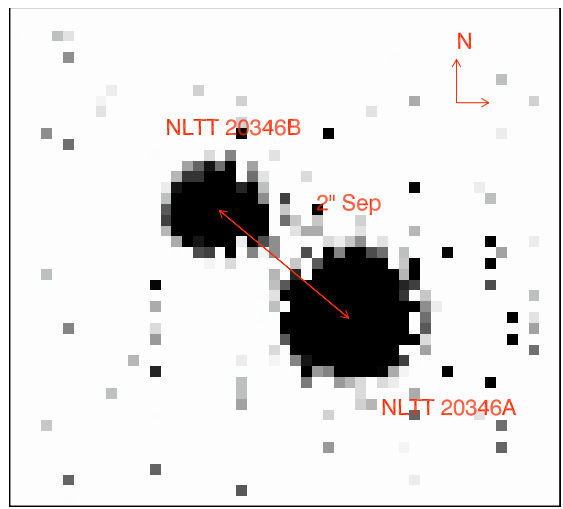}
\caption{ SpeX $J$ band image of the 6x6 arcsecond field around NLTT 20346 where we resolved the system into two M dwarf components.  The angular separation is $\sim$ 2 $\arcsec$ (or 50-80 AU) and the $J$ band magnitude difference is 1.13$\pm$0.02. } 
\label{fig:NLTT20346}
\end{figure}

In close examination of an ISPI image from 2009 November 30 (seeing conditions $\sim$0.7$\arcsec$), we noted an elongated point spread function (PSF) for NLTT 20346 in $\it{J}$ band after a 1s exposure indicating a visual binary.   Subsequent images taken with the $\it{Cont-203}$ filter (and $J,H,K$ filters on the SpeX guider camera--see section 2.3.2 below) resolved a companion separated by $\sim$2.0 $\arcsec$ (see Figure~\ref{fig:NLTT20346}).

\begin{figure}[!ht]
\begin{center}
\includegraphics[width=1.0\hsize]{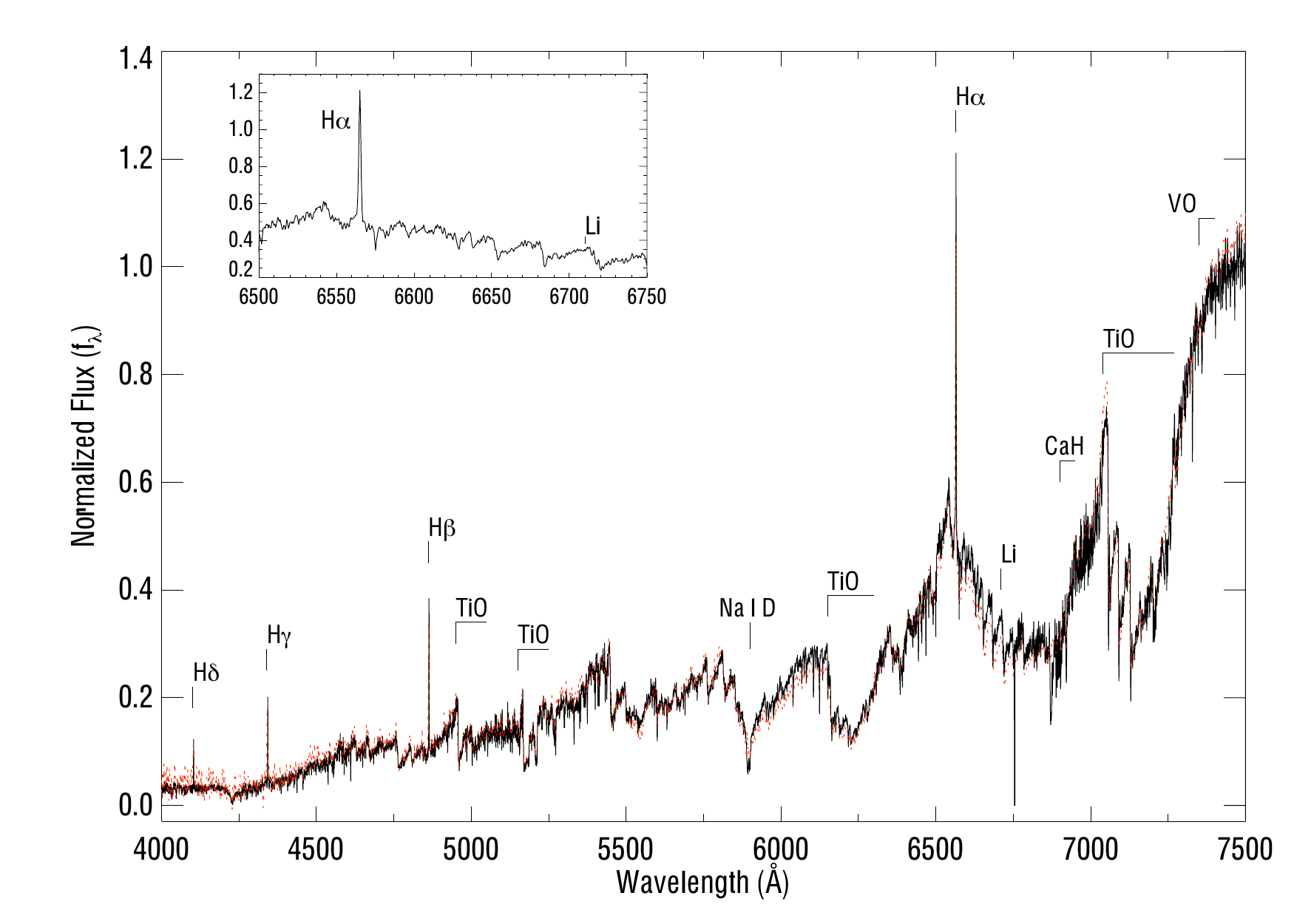}
\includegraphics[width=1.0\hsize]{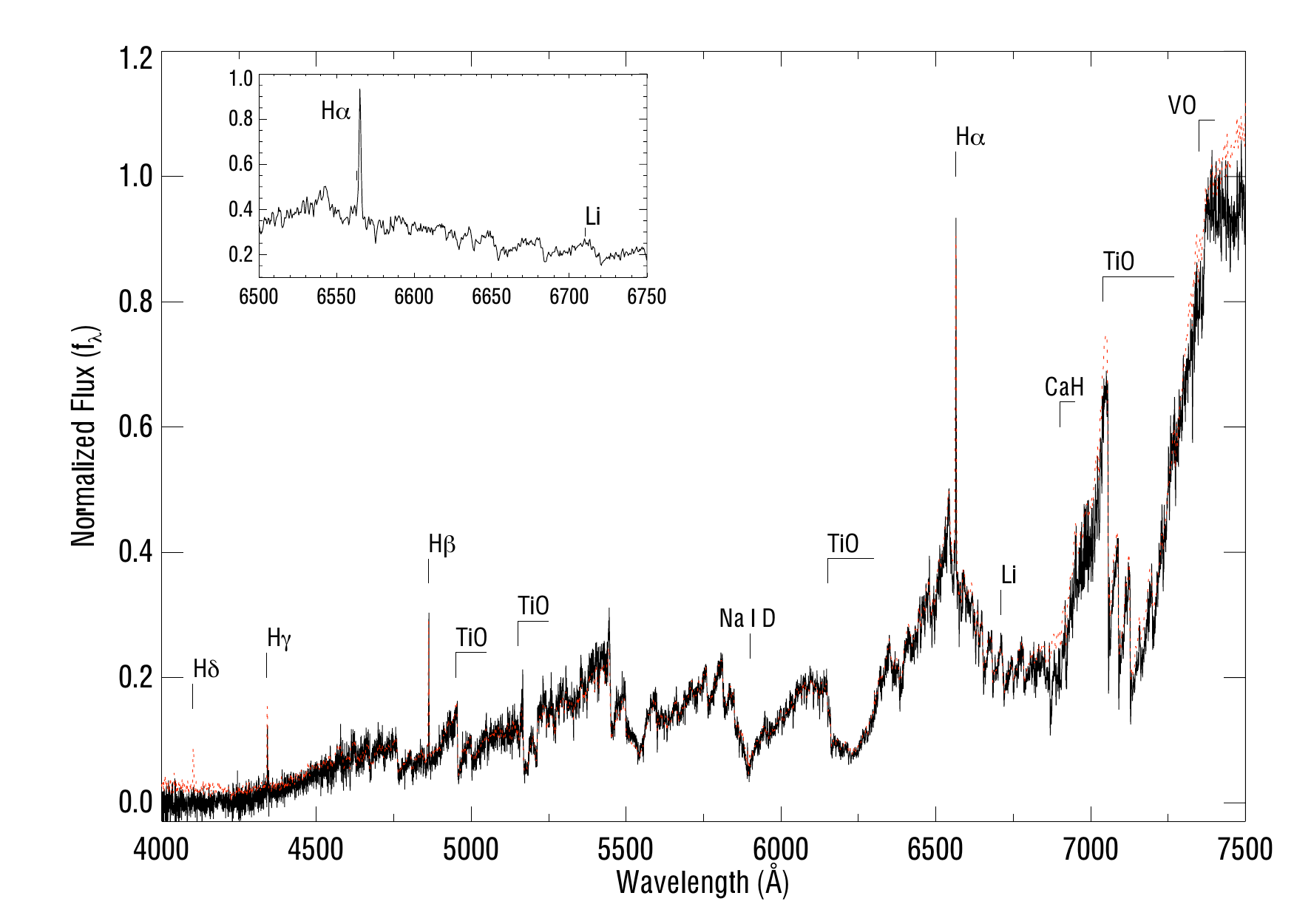}
\end{center}
\caption{ The optical spectrum using MagE of the primary (top plot) and secondary (bottom plot) components of the NLTT 20346 system.  Over-plotted is the template for an active M5 (top plot) and M6 (bottom plot) from \citet{2007AJ....133..531B} normalized at 7400~\AA~(dotted line).  The inset shows H$\alpha$ (6563~\AA) emission and a lack of Li (6708~\AA) absorption in both components. } 
\label{fig:NLTT20346_spectra}
\end{figure}

\begin{figure}[!ht]
\begin{center}
\includegraphics[width=1.0\hsize]{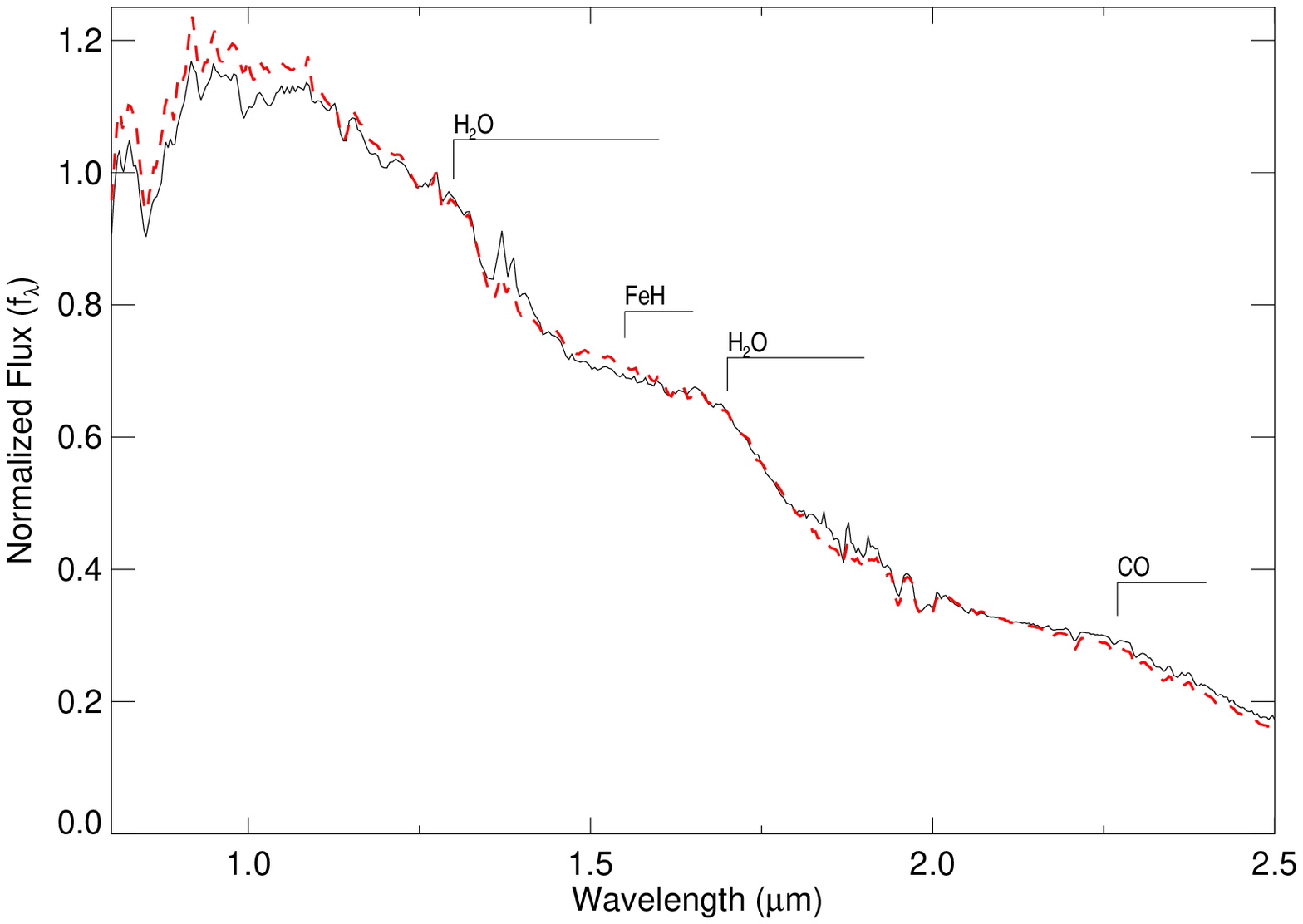}
\includegraphics[width=1.0\hsize]{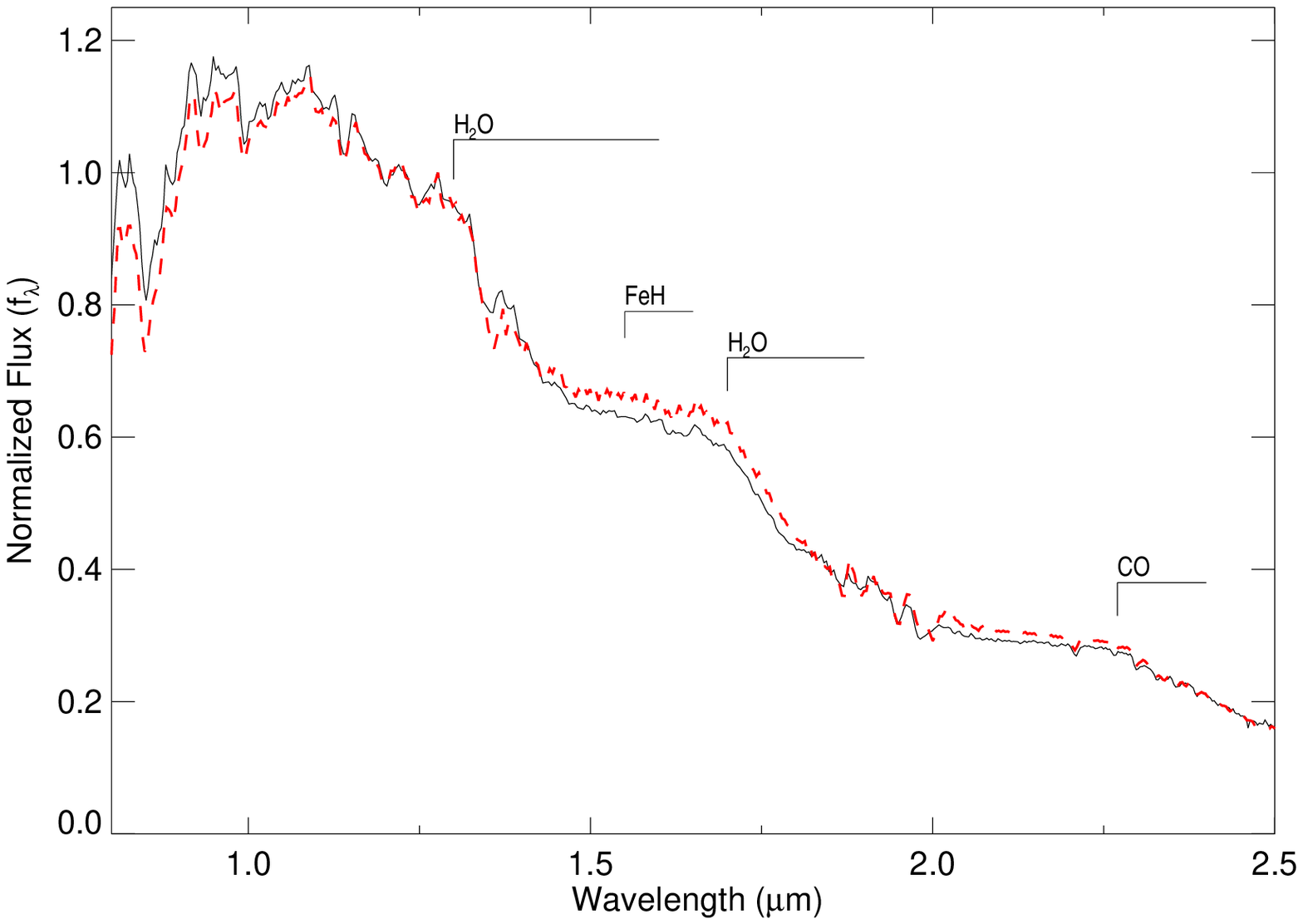}
\end{center}
\caption{ The near-IR spectrum using SpeX of the primary (top plot) and secondary (bottom plot) components of the NLTT 20346 system. Over-plotted as a dotted line is the M4 optical standard Gl 213 on the primary component (top) and the M5 optical standard Gl 51 on the secondary component (bottom). } 
\label{fig:near-IR}
\end{figure}

{\it \subsection{Spectroscopy}}
{\it \subsubsection{MagE}}
Optical spectra for the components of the NLTT 20346 visual binary were obtained with the Magellan Echellette Spectrograph (MagE; \citealt{2008SPIE.7014E.169M}) on the 6.5m Clay Telescope at Las Campanas Observatory on 2008 November 26 and 2010 March 9 (see Figure ~\ref{fig:NLTT20346_spectra}).  MagE is a cross--dispersed optical spectrograph, covering 3,000 to 10,000~\AA~ at medium resolution ($R \sim 4,100$).  Our observations employed a $0.7^{\prime\prime}$ slit aligned at the parallactic angle.   Observations were made under clear conditions with an average seeing of $\sim$0.7$\arcsec$.  The separation between the primary and secondary is $\sim$2$\arcsec$ therefore minimal contamination ($<$ 1\%) was expected.  A 300s integration was obtained for each component followed by a ThAr lamp spectrum for wavelength calibration.  The spectrophotometric standard GD 108 or EG 274 was observed for flux calibration. Ten Xe-flash and Quartz lamp flats as well as twilight flats were taken at the start of the evening for pixel response calibration.  The data were reduced using the MagE Spectral Extractor pipeline (MASE; \citealt{2009PASP..121.1409B}) which incorporates flat fielding, sky subtraction and flux calibration IDL routines. 


{\it \subsubsection{Near-Infrared Spectroscopy and Imaging with SPEX}}
We obtained low-resolution near-infrared (NIR) spectroscopy for both the primary and secondary components of NLTT 20346 using the SpeX spectrograph (\citealt{2003PASP..115..362R}) mounted on the 3m NASA Infrared Telescope Facility (IRTF) on 2009 December 3 UT (see Figure~\ref{fig:near-IR}) .  We used the spectrograph in prism mode with the 0$\farcs$5 slit aligned to the parallactic angle.  This resulted in $R~\equiv~\lambda$ / $\Delta\lambda~\approx$~120 spectral data over the wavelength range in 0.7--2.5 $\mu$m.  Conditions were clear and the seeing was 0$\farcs$46 at $K$.  The stars were separated sufficiently that we were able to obtain 12 individual exposure times of 30 seconds (total exposure time of 6  minutes) in an ABBA dither pattern along the slit.  Immediately after the science observations we observed the A0V star HD 43607 at a similar airmass for telluric corrections and flux calibration.  Internal flat-field and Ar arc lamp exposures were acquired for pixel response and wavelength calibration, respectively.  All data were reduced using the SpeXtool package version 3.4 (\citealt{2004PASP..116..362C}, \citealt{2003PASP..115..389V}) using standard settings.


In addition to NIR spectroscopy, we used the SpeX guider camera to image NLTT 20346 and resolve the two components with a S/N of $>$ 50 for each.  The resolution is 0$\farcs$12 per pixel with a 60$\arcsec$ $\times$ 60$\arcsec$ field of view.  We used the $JHK$ filters and the camera oriented with North up and East to the left.  Exposure times were 5 seconds each in an ABBA dither pattern, which was then shifted 10$\arcsec$ to the east and repeated (see Figure~\ref{fig:near-IR}).  A sky image was created from the 8 images then subtracted from individual frames before relative photometry was performed.  



\section{Astrometric Details}

{\it \subsection{Contaminant of 2MASS J0850+1057}}
2MASS J0850+1057 was targeted in the astrometric programs of \citet{2004AJ....127.2948V} and \citet{2002AJ....124.1170D} resulting in two published but discrepant parallax values.  The  \citet{2004AJ....127.2948V} work used a near-IR imager with either $J$ or $H$ filters ($H$ in the case of 2MASS J0850+1057) and the  \citet{2002AJ....124.1170D} work used an optical CCD with a $wide-I$ interference filter for astrometric measurements.  There were seven objects in common between the two programs, five of which had parallax measurements that matched better than 4 mas. 2MASS J0850+1057 was discrepant by 12.9$\pm$5.5 mas (CCD-IR) resulting in a 2.4$\sigma$ difference in parallactic angle and a 3.4$\sigma$ difference in proper motion position angle.    \citet{2004AJ....127.2948V} investigated whether orbital motion between the resolved components of the binary could account for the disagreement but ruled it out due to the short time-span between the two programs compared to the predicted $\sim$51 year orbit.   

\begin{figure}[htbp]
\begin{centering}
\includegraphics[width=1.0\hsize]{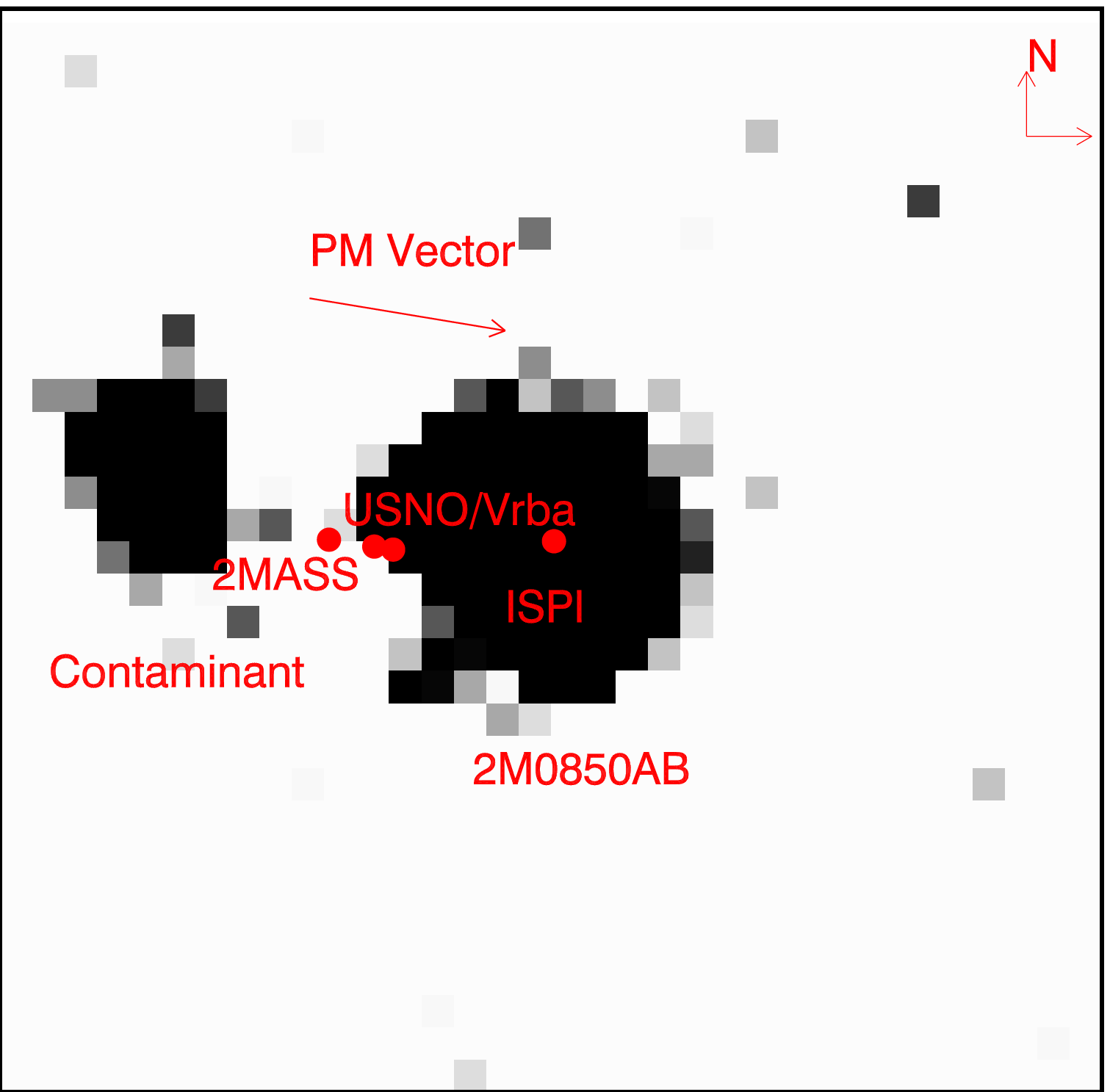}
\end{centering}
\caption{ ISPI $J$ band image of the 10x10 arcsecond field around the target 2MASS J0850+1057. The point source at the far left is $\sim$ 2 magnitudes fainter at $J$ but clearly resolved at the 2009 ISPI position.  Over-plotted with filled circles are the mean epoch positions from 2MASS, \citet{2002AJ....124.1170D}, and \citet{2004AJ....127.2948V}.  Tracing back the ISPI position to the \citet{2004AJ....127.2948V} and \citet{2002AJ....124.1170D} positions using the proper motion calculated in this work reveals that 2MASS J0850+1057 would have been blended with the contaminant. Over the lifetime of the two astrometric programs, the significant proper motion would lead to an "elongation" of the PSF which is a likely explanation for discrepant parallax measurements between the two programs. }
\label{fig:blend}
\end{figure}

We noticed on recent ISPI images that there was a second fainter object resolved in the $J$ band $\sim$ 4.0$\arcsec$ from the L dwarf that could have skewed prior astrometric measurements  (see Figure~\ref{fig:blend}).  The \citet{2004AJ....127.2948V} parallax measurement was based on 1.86 years of data, 13 images and a mean epoch of 2001.791.  The \citet{2002AJ....124.1170D} parallax measurement was based on 3.3 years of data, and 30 images.  Since no mean epoch is given, we assume observations were conducted between 1999-2002 shortly after the discovery and before the \citet{2002AJ....124.1170D} publication.  Both astrometric programs have comparable resolutions to ISPI.  Tracing the position of 2MASS J0850+1057 back to the \citet{2004AJ....127.2948V} and  \citet{2002AJ....124.1170D} mean positions from the ISPI position using our updated proper motion vector revealed that the binary L dwarf would have been blended with the contaminant at the time of the cited astrometric programs.  Over the 1.86-3.3 years of the programs, the significant proper motion of 2MASS J0850+1057 would have changed the centroid shape as its separation with the contaminant increased.  This potential "elongation" of the centroid would have skewed the position of 2MASS J0850+1057 in the frames used to calculate the parallax and proper motion.

A digital sky survey image from 1995 November 17 in $I$ band as well as a Sloan Digital Sky Survey (\citealt{2000AJ....120.1579Y}) image from 2005 November 06 in the sloan $i$ band also show the point source.  The second object appears to be 2 magnitudes fainter in $J$ but shows no appreciable motion and we conclude that it is unrelated to the brown dwarf.    We propose that this contaminating source is the most probable cause for the discrepant parallax results.  



\begin{deluxetable*}{lccccccrrrrrrrr}
\label{tab:tab1}
\tabletypesize{\scriptsize}
\tablecaption{Details on Astrometry\label{PM}}
\tablewidth{0pt}
\tablehead{
\colhead{Name} &
\colhead{RA} &
\colhead{DEC} &
\colhead{$\mu_{\alpha}$ } &
\colhead{$ \mu_{\delta}$}  &
\colhead{$v_{rad}$}  &
\colhead{Distance} &
\colhead{Reference} \\
 &
 \colhead{Epoch 2000} &
\colhead{Epoch 2000} &
\colhead{\arcsec yr$^{-1}$} &
\colhead{\arcsec yr$^{-1}$}&
\colhead{km s$^{-1}$} &
\colhead{pc}\\
\colhead{(1)}  &
\colhead{(2)}  &
\colhead{(3)}  &
\colhead{(4)}  &
\colhead{(5)}  &
\colhead{(6)}  &
\colhead{(7)}  &
\colhead{(8)}  \\}

\startdata
NLTT 20346\tablenotemark{a}		&08 50 19.18&+10 56  43.69&	-0.172 $\pm$ 0.008\tablenotemark{b}	&-0.050 $\pm$ 0.006\tablenotemark{b}	&	26 $\pm$ 9\tablenotemark{c}&  31 $\pm$ 7\tablenotemark{d}   &1\\	
NLTT 20346						&---&---&	-0.179 $\pm$ 0.010	&-0.052 $\pm$ 0.010&	---&---&2\\	
NLTT 20346						&---&---&	-0.177 $\pm$ 0.019	&-0.035 $\pm$ 0.019&	---&---&3\\	
\hline
NLTT 20346A	&---&---&	-0.178 $\pm$ 0.033\tablenotemark{e} &-0.050 $\pm$ 0.032\tablenotemark{e}  &25 $\pm$ 7   & 28 $\pm$ 5&1\\
NLTT 20346B	&---&---&	-0.166 $\pm$ 0.033\tablenotemark{e} &-0.050 $\pm$ 0.032\tablenotemark{e} &26 $\pm$ 7   & 33 $\pm$ 5&1\\
\hline
2MASS J0850+1057&08 50 35.93 & +10 57 15.62&	-0.144 $\pm$ 0.006	&-0.038 $\pm$ 0.006	&	---&29 $\pm$ 7&1\\	
2MASS J0850+1057&---&---&-0.143 $\pm$ 0.006	&-0.020 $\pm$ 0.003	&	---&38 $\pm$ 6   &4\\	
2MASS J0850+1057&---&---&-0.142 $\pm$ 0.002	&-0.008 $\pm$ 0.002	&	---&25.6 $\pm$ 2.3  &5\\	

\hline
\enddata
\tablerefs{Refs--1=This paper 2=\citet{2002AJ....124.1190L} 3=\citet{1979lccs.book.....L} 4=\citet{2004AJ....127.2948V} 5=\citet{2002AJ....124.1170D} } 
\tablenotetext{a}{NLTT 20346 consists of two M dwarf components separated by $\sim$2$\arcsec$ or 50 - 80 AU.}
\tablenotetext{b}{Based on the catalog positions of the blended combined light source over a 59 year baseline}
\tablenotetext{c}{Based on the mean RV value of NLTT 20346A and B.}
\tablenotetext{d}{Based on the mean value from the spectrophotometric distance to the components. }
\tablenotetext{e}{Based on the positions from resolved SDSS and ISPI  images with a $\sim$4 year baseline.}
\end{deluxetable*}

{\it \subsection{Proper Motion}}
2MASS J0850+1057 was imaged as part of the Brown Dwarf Kinematics Project  (BDKP) parallax program which targets nearby brown dwarfs.  We used the Carnegie Astrometric Planet Search software (from here-on ATPa) to extract all point sources in the 6 epochs of ISPI data and solve for the parallax and proper motion (\citealt{2009PASP..121.1218B}).  The highest quality image was used as the template which all other images were transformed.   In the 2MASS J0850+1057 field, there were 25 well-behaved (elongated, saturated, spurious sources were removed)  reference stars between all epochs.  Using these sources, a linear transformation was applied to each epoch point source catalog to constrain the field rotation, plate scale and match all reference sources to the template.  Higher order transformations were tested but demonstrated negligible difference from linear solutions.  The apparent trajectory of each star was fit to a standard astrometric model included in the ATPa software.  The algorithms follow the astrometric solution prescriptions laid out in the Hipparcos (\citealt{1997A&A...323L..49P}) and Tycho Catalogue (\citealt{2000yCat.1259....0H}) descriptions\footnote{$http://www.rssd.esa.int/SA/HIPPARCOS/docs/vol1_all.pdf$}.   We corrected the data from apparent parallax to absolute parallax following the same procedure described in \citet{2004AJ....127.2948V}.  A full detailed description of the astrometric pipeline used to solve for the parallax and proper motion of 2MASS J0850+1057 is provided in Faherty et al. (in prep).  As noted in Vrba et al (2004) the proper motion and parallax of the system will not be effected by the orbital motion since the images were taken over a $\sim$ 2 year period whereas the orbit is $\sim$ 50 years with components separated by 0.16$\arcsec$.

To determine the new proper motion for NLTT 20346 we used positions available through USNO-A2.0, USNO-B1.0, GSC 2.2, 2MASS, and an ISPI image taken on 08 April 2010 (\citealt{2003AJ....125..984M}, \citealt{1998usno.book.....M};\citealt{2008AJ....136..735L}; \citealt{2003tmc..book.....C}).   The total baseline between the first and final epoch used was $\sim$59 years.  NLTT 20346 is not resolved in any of the epochs as the 2$\arcsec$ separation, seeing conditions and plate scale of each detector caused the A and B components to appear blended.  We simulated the PSF for the A and B component of NLTT 20346 on each of the detectors to estimate the uncertainty due to blending.  We added this additional uncertainty in quadrature to the catalog uncertainties and used these in the proper motion measurement.  The absolute astrometric position from each catalog (with corresponding epochs) were used to solve for the proper motion using a least squares weighted solution.  There is an SDSS image taken on 06 January 2006 and an ISPI image taken on 27 February 2010 with sub-arcsecond seeing where both components are resolved.  Using these two epochs with a $\sim$4 year baseline, we calculated the proper motion of each component.  These resolved values are consistent with the motion calculated from the blended source along a 59 year baseline.  This indicates that orbital motion is not effecting the total motion of the system.  

For NLTT 20346 we calculated $\mu_{\alpha}$=-172$\pm$8 mas yr$^{-1}$ and $\mu_{\delta}$=-50$\pm$6 mas yr$^{-1}$ (using all catalog positions) and for  2MASS J0850+1057 we calculated $\mu_{\alpha}$=-144$\pm$6 mas yr$^{-1}$ and $\mu_{\delta}$=-38$\pm$6 mas yr$^{-1}$.  NLTT 20346 has two published proper motion values in the LSPM-N and New Luyten (NLTT--\citealt{1979lccs.book.....L}) catalogues and, as stated in section 3.1, 2MASS  J0850+1057 has previous proper motion values reported in \citet{2004AJ....127.2948V} and \citet{2002AJ....124.1170D} (see Table ~\ref{PM}).  Our astrometric results are consistent within 2$\sigma$ of previous published results and using the new values , both the $\mu_{\alpha}$ and $\mu_{\delta}$ values for the potential companions are within 2$\sigma$ of each other.  Along with the predicted distance measurement of the primary (see section 4.1 below), and the new parallax measurement for 2MASS J0850+1057 of 35$\pm$8 mas or 29$\pm$7 pc, this system is a strong wide companion common proper motion candidate.


{\it \subsection{Likelihood of Companionship}}
NLTT 20346 was identified as a potential companion to 2MASS J0850+1057 through a common proper motion search of the Brown Dwarf Kinematics Project  (BDKP) catalog (\citealt{2009AJ....137....1F}) and the Lepine-Shara Proper Motion North (LSPM-N) and Hipparcos catalogs (\citealt{2002AJ....124.1190L};  \citealt{1997A&A...323L..49P}).   Several other systems were detected and detailed analysis was presented in \citet{2009AJ....137....1F}.  In that work, an angular separation of up to 10 arcminutes and a proper motion match criterion of better than 2$\sigma$ in both right ascension (RA) and declination (DEC) between the system components was required to determine common proper motion candidates. The average uncertainty for objects in the BDKP catalog is 15 mas yr$^{-1}$ (in both directions) so proper motion agreement was typically required to be $<$ 30 mas yr$^{-1}$ between the stellar companion and ultracool dwarf (UCD). \citet{2010AJ....139..176F} also required a distance match between components of better than 2$\sigma$ or typically better than 10 pc. 

The system containing NLTT 20346 was not investigated in \citet{2010AJ....139..176F} because proper motion components were slightly outside of the 2$\sigma$ requirement.  However, after follow-up imaging and re-analysis of the astrometry of both components we found strong evidence for companionship (see Section 3.2).	

To quantify the probability that NLTT 20346 might be a chance alignment with 2MASS J0850+1057, we ran a Monte Carlo simulation of all stars in the LSPM-N and Hipparcos catalogs that shared a common proper motion, but not necessarily distance or position, with the brown dwarf (to within 2$\sigma$--see \citealt{2009AJ....137....1F} for details).  There were 156 stars in the Hipparcos catalog and 632 stars in the LSPM-N catalog with matching proper motion components. After 10000 iterations we found the likelihood that NLTT 20346 is a chance coincidence with 2MASS J0850+1057 (at an angular seperation of 248 $\arcsec$)  is  $<$ 1.0\%.

\section{SYSTEM PARAMETERS}
{\it \subsection{Binary Components of NLTT 20346}}

From the MagE data for NLTT 20346 we classified the brighter component as an M5$\pm$0.5 and the fainter as an M6$\pm$0.5 by comparing to active M dwarf templates (see Figure ~\ref{fig:NLTT20346_spectra}) in \citet{2007AJ....133..531B}.  SpeX imaging of NLTT 20346 allowed us to measure the relative magnitude difference between components ($\Delta$$J$=1.13$\pm$0.02) and subsequent $J$ band magnitudes of 11.61$\pm$ 0.04 and 12.74$\pm$ 0.08 for the M5 and M6 respectively.  From the SpeX spectrum, we classify the brighter of the two components as an M4$\pm$1 and the fainter as an M5$\pm$1 by comparing to the near-IR spectra of the M4 optical standard Gl 213 and the M5 optical standard Gl 51 respectively (see Figure~\ref{fig:near-IR}).  This is consistent within uncertainties with the optical spectral types.  In the absence of a parallax measurement we computed a spectrophotometric distance to each component using the spectral type, individual $J$ band magnitudes, and the spectrophotometric $J$ band relation from Golimowski et al (2010 in prep). We calculated component distances of 28$\pm$5 pc and 33 $\pm$ 5 for the M5 and M6 respectively resulting in a mean distance value to NLTT 20346 of 31$\pm$7 pc.  



\begin{figure}[!ht]
\begin{center}
\epsscale{2.0}
\plottwo{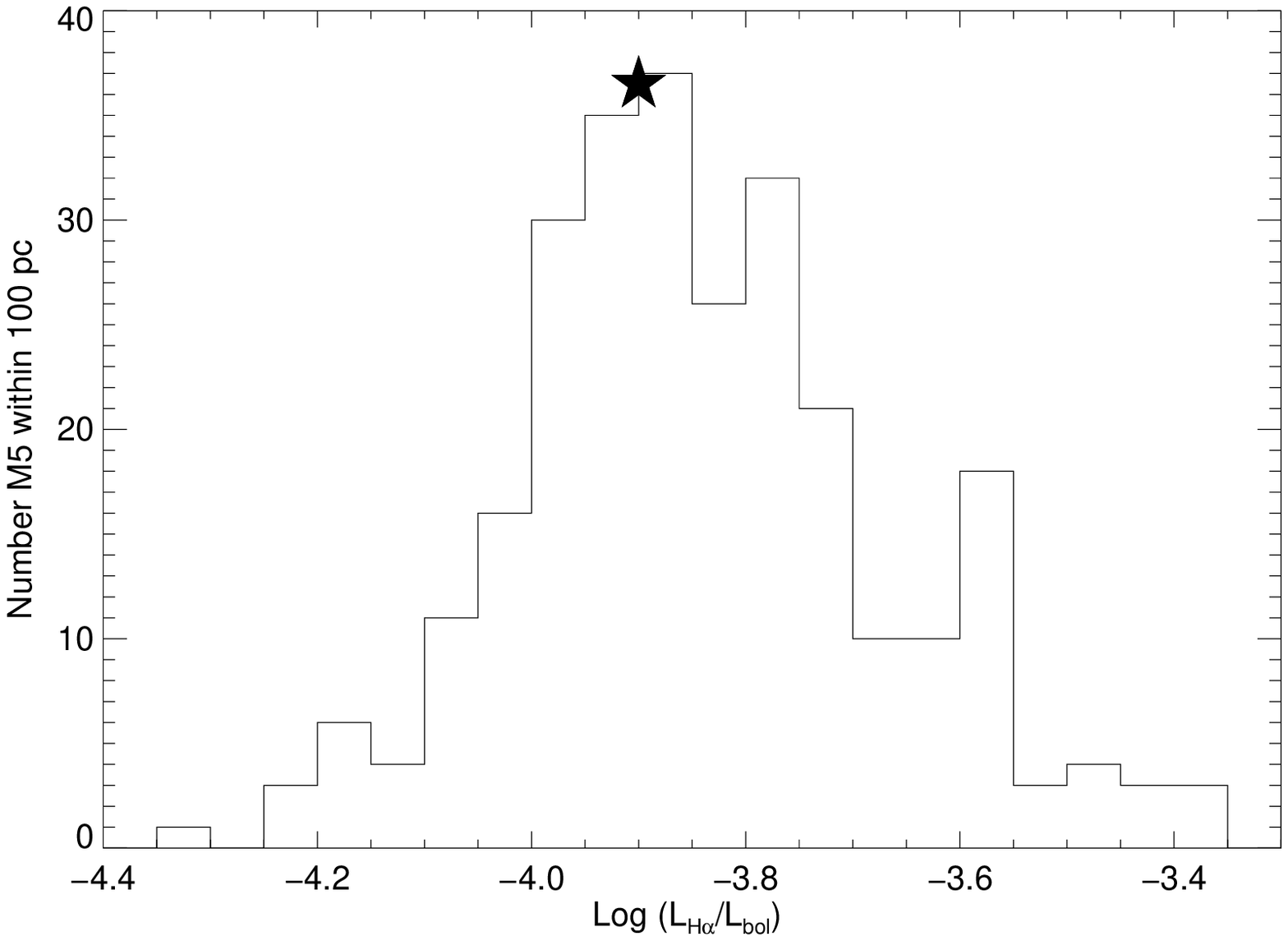}{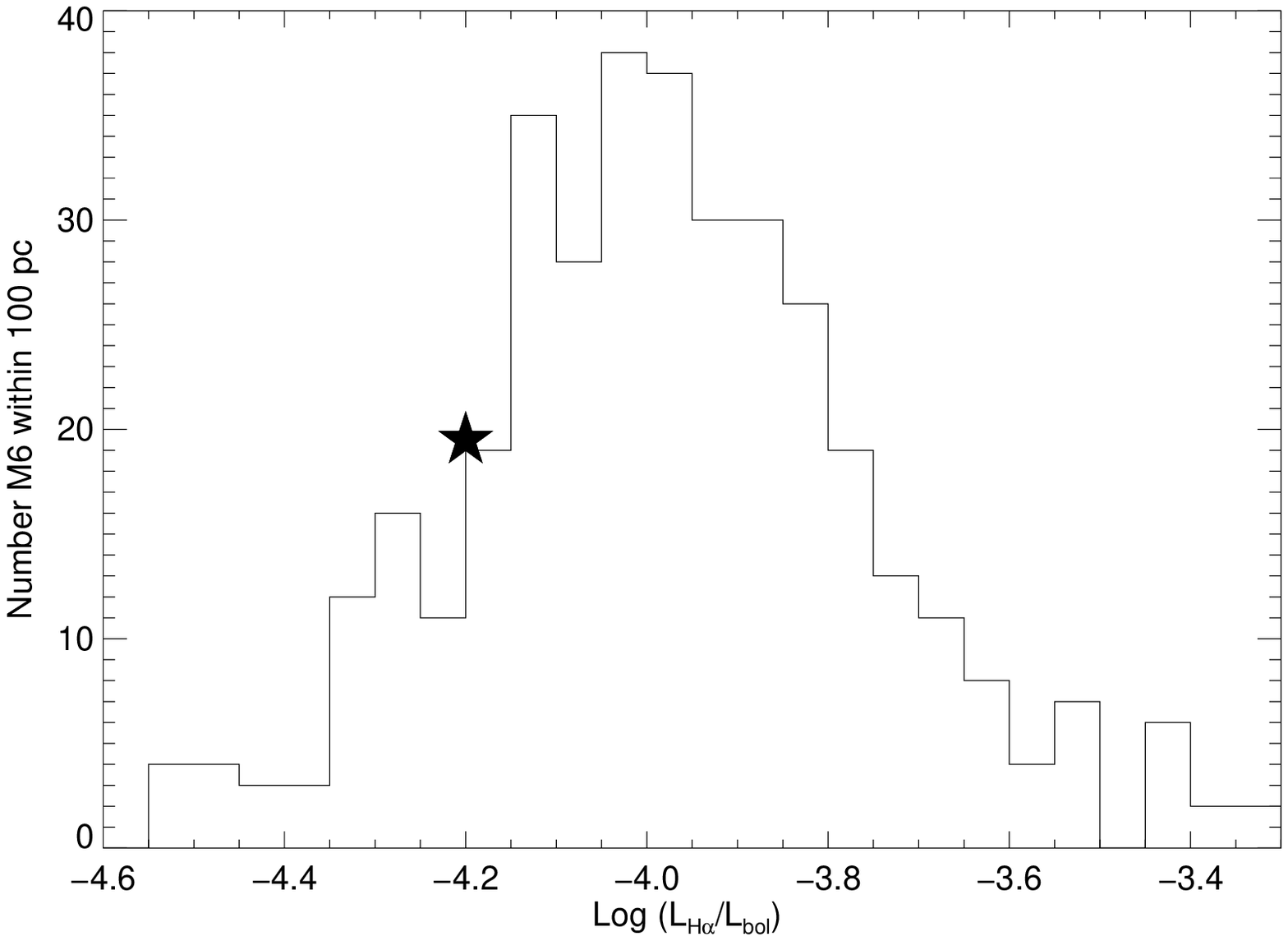}
\end{center}
\caption{Distribution of log(L$_{H\alpha}$/L$_{bol}$) for active M5 (top plot) and M6 dwarfs (bottom plot) in the \citet{2008AJ....135..785W} spectroscopic sample within 100 pc of the Sun.  NLTT 20346A (top) and NLTT 20346B (bottom) have values of -3.94$\pm$0.02 and -4.20$\pm$0.02 respectively (marked by a five point star), within 1$\sigma$ of typical active objects at similar temperatures.} 
\label{fig:Halpha}
\end{figure}

{\it \subsection{Activity and Age of NLTT 20346}}
The optical spectra of both components of NLTT 20346 show moderate H$\alpha$ emission.  We measured an H$\alpha$ equivalent width of 4.20$\pm$0.06~\AA~ and  3.64$\pm$0.08~\AA~ from the MagE data for the M5 and M6, respectively.  The upper Balmer series showing H$\beta$ through H$\delta$ as well as Ca II K and H+H$\epsilon$ are also seen in emission in both components (see Figure ~\ref{fig:NLTT20346_spectra}).   Combining the H$\alpha$ equivalent width with the $\chi$ parameter from \citet{2004PASP..116.1105W} yields the log(L$_{H\alpha}$/L$_{bol}$), a metric of magnetic activity and a statistical proxy for age.  \citet{2008AJ....135..785W} have found mean log(L$_{H\alpha}$/L$_{bol}$) values for active M5 and M6 dwarfs within 100 pc of the Sun (see Figure ~\ref{fig:Halpha}) of -3.9$\pm$0.2 and -4.0$\pm$0.3, respectively. NLTT 20346A and B have  log(L$_{H\alpha}$/L$_{bol}$) values of -3.94$\pm$0.02  and -4.20$\pm$0.02 respectively, both within 1$\sigma$ of typical active M dwarfs.  \citet{2008AJ....135..785W} estimate an activity lifetime for M5 and M6 dwarfs of 7.0$\pm$0.5 Gyr.    Using the age activity relation from \citet{2009IAUS..258..327W}\footnote{Equation 3.1 in \citet{2009IAUS..258..327W} contains an error.  The functional form should be log(L$_{H\alpha}$/L$_{bol}$)=$\frac{-a}{l^{n}-t^{n}}-b$ } yields consistent component ages of 6.3$\pm$1.0 and 6.5$\pm$1.0 Gyr for the M5 and M6 respectively.  

\begin{figure}[htbp]
\begin{centering}
\includegraphics[width=1.0\hsize]{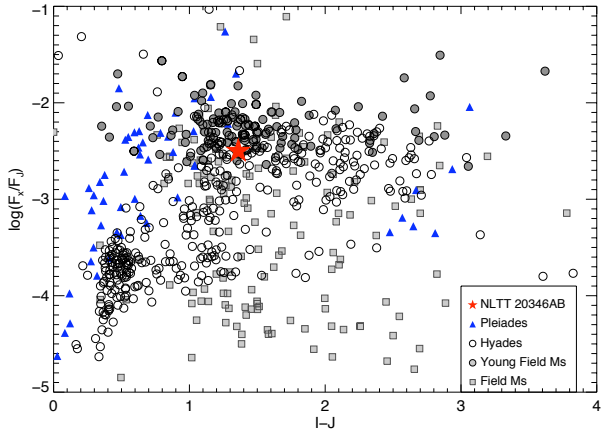}
\end{centering}
\caption{ Comparison of the X-ray activity of NLTT 20346 to M dwarfs in the Pleiades, Hyades, young and older field M dwarfs (data from \citealt{2009ApJ...699..649S}). NLTT 20346 is marked by the five point star.  Note that the log (f$_{x}$/f$_{J}$) for NLTT 20346 is based on the combined light in $J$ band and X-ray as we do not have resolved X-ray data for the components.} 
\label{fig:xray}
\end{figure}

The NLTT 20346 system is an X-ray source and the X-ray activity indicates a younger age than that calculated from the H$\alpha$ activity.  2XMMiJ085018.9+105644 (from the X-ray Multi Mirror Mission(XMM)-Newton 2nd Incremental Source Catalogue) is 2$\arcsec$ from the position of the source.  We computed the X-ray flux (f$_{X}$) by combining the 0.2-0.5 keV, 0.5-1.0 keV, and 1.0-2.0 keV ranges  in order to compare to equivalent X-ray detections of M dwarfs in the R{\"o}ntgen Satellite (ROSAT; \citealt{1999A&A...349..389V}) catalog ($\sim$ 0.2-2.4 keV).   We used the f$_{x}$/f$_{J}$ relation defined in \citet{2009ApJS..181..444A}:  log (f$_{x}$/f$_{J}$)=log($f_{x}$)+0.4$J$+6.30 to compute  log (f$_{x}$/f$_{J}$) for NLTT 20346 of -2.5. Note that the 2MASS $J$ band magnitude for the combined light of the system was used to calculate log (f$_{x}$/f$_{J}$) since the components were not resolved in the XMM-Newton data.  Figure ~\ref{fig:xray} compares this estimate to Hyades, Pleiades, Young ($<$ 300 Myr) and thin-disk field M stars (\citealt{2009ApJ...699..649S}).  The objects labeled Field Ms are from the \citet{1999A&AS..135..319H} study.  Several of these objects demonstrate large X-ray flux although they are regarded in \citet{2009ApJ...699..649S} as normal field objects (typically $>$ 1 Gyr). Possible explanations include flaring M dwarfs (\citealt{2010AJ....140.1402H} calculate a flare duty cycle for late M dwarfs of as much as 3\%), or undetected spectroscopic binaries (\citealt{2009ApJ...699..649S} found a 16\% contamination rate of SBs in their sample).  

\citet{2005ApJS..160..390P} investigated the age-activity trend for low mass stars.  Comparing X-ray activity for stars in the mass range  0.1$<$M$_{\sun}$ $<$0.4 belonging to nearby associations or clusters with ages spanning 1-100 Myr (comparison samples were ONC, NGC 2264, Chamaeleon, Pleiades, and Hyades) they found only a mild decrease in X-ray luminosity over the 1-100 Myr range.  However for the same mass range they found that significant decay was detected over longer timescales (such as from Hyades age objects to Field--several Gyr-- age objects).  Unfortunately no relation is derived for obtaining an age for the slightly older population of M dwarfs based on X-ray luminosity.  On  Figure ~\ref{fig:xray}, the X-ray activity of NLTT 20346 places it with the most active Hyades stars or the least active Pleiades stars showing similar colors.  It also has X-ray activity levels comparable to the young M dwarf sample from \citet{2009ApJ...699..649S} where the estimated age of the sample was $<$ 300 Myr (with objects well above the X-ray activity level of the Hyades having ages $<$ 150 Myr).  To estimate the age from the X-ray activity level we assume a Skumanich like decay from the mean log (f$_{x}$/f$_{J}$) of the Pleiades (-2.3 $\pm$0.3) with an assumed age of $\sim$ 120 Myr.  The resulting age range for NLTT 20346 is 250-450 Myr consistent with the young field M dwarf sample for which it has comparable activity levels.  We note that there are a handful of M dwarfs in the \citet{1999A&AS..135..319H} sample which show comparable activity levels indicating that we cannot rule out that NLTT 20346 is a flare star or spectroscopic binary (explanations that could account for the high X-ray levels).


The $\sim$2$\arcsec$ or 50-80 AU separation of the components of NLTT 20346 is safely outside the distance where tidal interactions would affect stellar rotation and hence the chromospheric age of the system (\citealt{2007ApJ...665L.155M}).  However, \citet{1996MNRAS.280..458A} find that binary systems with separations $<$100 AU can suffer accelerated mass-accretion from star-disk interactions early in their lifetimes which could lead to variations in the rotation rate.  The NLTT 20346 system is close to the border where disk interactions become significant, so the X-ray age above might be skewed to a slightly younger age.   


We cross-correlated the MagE spectra of NLTT 20346A and B with the M5 and M6 templates from \citet{2007AJ....133..531B} using the $xcorl$ package in IDL (\citealt{2003ApJ...583..451M}; \citealt{2009ApJ...693.1283W}) to measure consistent radial velocities of  25$\pm$7 km/s and 26$\pm$7 km/s respectively.  Combining the mean RV value of 26$\pm$9 km/s with the proper motion, position, and spectro-photometric distance leads to (U,V,W) values of (21$\pm$5,-10$\pm$4,-1$\pm$4) km/s (corrected for solar motion with U positive toward the Galactic center).  While, individual space motions can not be used to date objects, the calculated Galactic space motion for NLTT 20346 is  consistent with thin disk membership (age $<$ $\sim$3 Gyr), in agreement with the chromospheric diagnostics (\citealt{1989AJ.....97..431E,1989PASP..101...54E}).

Discrepancies between various age diagnostics are not surprising.  Previous studies have shown similar inconsistencies when determining ages for individual objects using different techniques (e.g. isochronal ages, gyrochronolgy, chromospehric activity, Lithium abundances; see \citealt{2010ARA&A..48..581S}).  Some possible explanations for the differing age values of this system are that (1) we could have observed the components of NLTT 20346 when the system was in a low state as variations of more than 30\% have been seen in H$\alpha$ measurements of M dwarfs (e.g. \citealt{2009ApJ...695..310B}).  (2) Preibisch \& Feigelson (2005) show the X-ray levels of objects $<$ 500 Myr versus field age objects with median ages of a few Gyr; however very little is known of what to expect for intermediary or juvenile age objects.  In this work we favor the younger age for the system as this shows consistency with the presence of strong Lithium absorption in 2MASS J0850+1057(see section ~\ref{0850}), the X-ray activity in NLTT 20346, and the kinematics of the system.



{\it \subsection{Age of 2MASS J0850+1057\label{0850}} } 
The optical spectrum of 2MASS J0850+1057 from \citet{2001AJ....121..489R} exhibits deep Li absorption (15.2 \AA~equivalent width) indicating component masses below $\sim$0.06 M$_{\sun}$ and an upper age limit between 0.80 -- 1.5 Gyr (\citealt{1992ApJ...389L..83R}; \citealt{1993ApJ...404L..17M}).  The strong Li absorption is consistent with the X-ray activity and subsequent younger age of NLTT 20346.  Coupled with the matching proper motion components and distance measurements further confirms the likelihood of companionship.  

We choose two age ranges for further analysis of this system: (1) 0.25 - 1.5 Gyr based on a lower limit constrained by the X-ray activity of the primary and an upper limit based on the Li detection in the secondary  and (2) 5 - 7 Gyr based on the H$\alpha$ activity of the primary.

\begin{figure}[htbp]
\begin{centering}
\includegraphics[width=1.0\hsize]{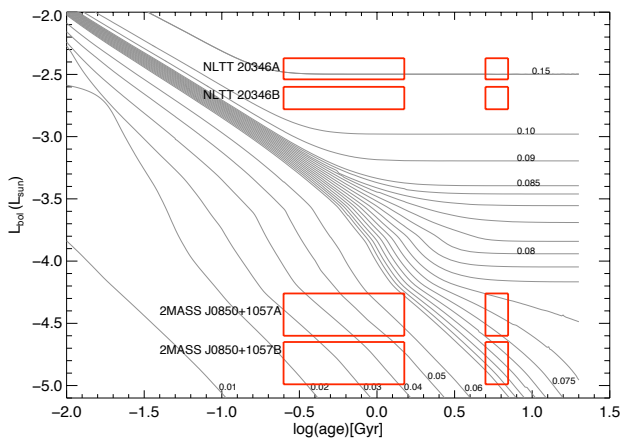}
\end{centering}
\caption{A plot of the \citet{1997ASPC..119....9B} evolutionary models with parameters (age and L$_{bol}$) for both the primary and secondary of each component of the quadruple system indicated with labeled boxes.  Masses are shown derived for the favored age range of 0.25 - 1.5 Gyr as well as the H$\alpha$ age range of 5 - 7 Gyr. } 
\label{fig:burrows}
\end{figure}

{\it \subsection{Component Masses}}
We determined masses for all components of the new system using the two age ranges discussed above (0.25-1.5 Gyr and 5 - 7 Gyr), L$_{Bol}$ values (described in this section), and the \citet{1997ASPC..119....9B} evolutionary models (see Figure~\ref{fig:burrows}) .   To compute L$_{bol}$ values for 2MASS J0850+1057 we combined our new parallax measurement with the resolved near-IR photometry and spectral type components reported in Burgasser et al (2010), and the bolometric correction calculated from \citet{2009ApJ...704.1519D}.  We determine log(L$_{bol}$/L$_{sun}$) values of -4.43$\pm$0.20 and -4.82$\pm$0.20 for the primary and secondary respectively.  Using evolutionary models and the younger age range of 0.25-1.5 Gyr, the calculated component masses for 2MASS J0850+1057 are 0.04$\pm$0.02 M$_{\sun}$ and 0.035$\pm$0.015 M$_{\sun}$ yielding a total brown dwarf binary mass (M$_{tot}$) of 0.075$\pm$ 0.025 M$_{\sun}$.  Using the older age of 5 - 7 Gyr the component masses are significantly higher; 0.076$\pm$0.02 M$_{\sun}$ and 0.072$\pm$0.03 M$_{\sun}$  for the primary and secondary respectively yielding a total brown dwarf binary mass (M$_{tot}$) of 0.148$\pm$ 0.040 M$_{\sun}$.  Table ~\ref{components} lists all system parameters. 

\begin{deluxetable*}{lccccccrrrrrrrr}
\label{tab:tab1}
\tabletypesize{\scriptsize}
\tablecaption{Details on System Components\label{components}}
\tablewidth{0pt}
\tablehead{
\colhead{Parameter} &
\colhead{ 2MASS J0850+1057} &
\colhead{NLTT 20346} &
\colhead{Reference}\\
}
\startdata
Combined Spectral Type						&L6					& M5					&1,2,5\\
Est. Component Types				&L7+L6.5				& M5+M6				&3,5\\
2MASS J							& 16.465 $\pm$ 0.113	& 11.282$\pm$0.023	&4\\
2MASS K							& 14.473$\pm$ 0.066	& 10.407$\pm$0.020	&4\\
$\Delta$ $J_{MKO}$						&	---				&1.13$\pm$0.02		&5\\
$\Delta$ $K_{MKO}$					&0.95$\pm$0.02		&---					&3\\
Primary log(L$_{bol}$/L$_{\sun}$)\tablenotemark{a}		&-4.43$\pm$0.20 		&-2.46$\pm$0.09		&5,7\\
Secondary log(L$_{bol}$/L$_{\sun}$)	& -4.82$\pm$0.20		&-2.69$\pm$0.09		&5,7\\
Primary Mass (M$_{\sun}$)\tablenotemark{a}			& 0.04$\pm$0.02 		&  0.16 $\pm$0.01		&5\\
Secondary Mass (M$_{\sun}$)			& 0.035$\pm$0.015		& 0.13$\pm$0.01 		&5\\
System Age (Gyr)				& 0.25-1.5				& 0.25-1.5				&5\\

\enddata
\tablerefs{Refs--1= \citet{1999ApJ...519..802K}; 2=\citet{2000AJ....120..447K};3=Burgasser et al (2010); 4= \citet{2003tmc..book.....C}; 5=This paper; 6=\citet{2010arXiv1001.4800K}; 7=\citet{2005nlds.book.....R}} 
\tablenotetext{a}{Assuming this component is a single; see Burgasser et al. (2010).}
\end{deluxetable*}

 \citet{2010arXiv1001.4800K}  calculated the total mass of this system to be 0.2$\pm$ 0.2 M$_{\sun}$ based on Keck AO observations and orbital fitting.   The large uncertainty of the dynamical mass measurement precludes a useful comparison with our result; however, they also calculate the mass from the evolutionary models of \citet{1997ASPC..119....9B} and \citet{2000ApJ...542..464C}. Based on an interpolation over a grid of temperature, luminosity, mass, and age provided by the evolutionary models (the Tucson models from  \citealt{1997ASPC..119....9B} and the DUSTY models from \citealt{2000ApJ...542..464C}) they conclude M$_{tot}$ to be 0.08$\pm$0.06, consistent with our derived value from the younger age range.   However, \citet{2010arXiv1001.4800K}  also conclude that objects of spectral type late M through mid L tend to have their masses under-predicted by current evolutionary models (see also \citealt{2008arXiv0807.2450D}).  With refined uncertainties on the mass in combination with the age information that can be gathered from NLTT 20346, 2MASS J0850+1057 will be a useful system for testing this result.

 
 We also calculated the component masses of NLTT 20346 using the evolutionary tracks of  \citet{1997ASPC..119....9B}.  Using the L$_{bol}$ values for an M5 and M6 from \citet{2005nlds.book.....R} combined with the age described above, yields component masses of 0.16 $\pm$0.01 M$_{\sun}$ and  0.13$\pm$0.01 M$_{\sun}$ or a total mass for NLTT 20346 of 0.29$\pm$0.01 M$_{\sun}$.   This is in agreement with \citet{2000A&A...364..217D} where the masses of an M5 and M6 are calculated to be 0.14$\pm$0.1 and 0.11$\pm$0.1(assuming a 10\% uncertainty), respectively, using a fourth degree polynomial fit to parallax data.
 
 Combining the mass results for 2MASS J0850+1057 using the younger age range with those of NLTT 20346 yields a total mass for this quadruple of 0.36 $\pm$0.02 M$_{\sun}$.

\section{DISCUSSION}

\begin{figure}[htbp]
\begin{center}
\includegraphics[width=1.0\hsize]{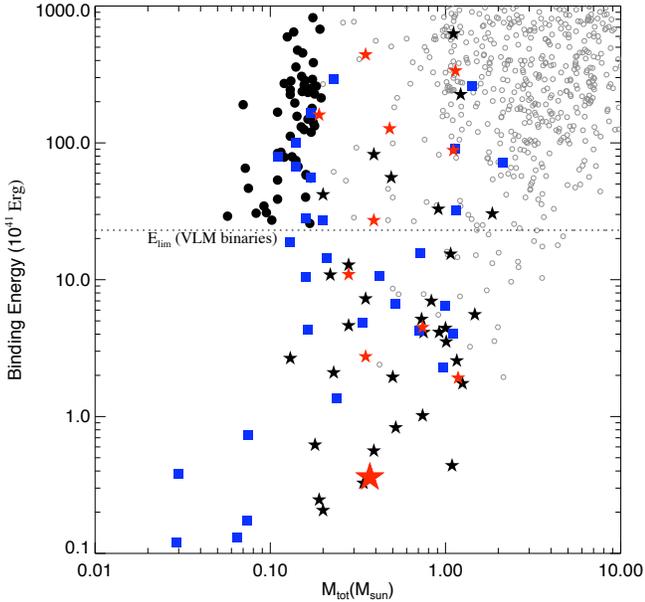}
\end{center}
\caption{A plot of the binding energy vs. total mass. Objects marked with filled circles are tight low--mass systems (typically M$_{tot}$ $<$ 0.2M$_{\sun}$ and $\rho<$20 AU). Wide systems ($\rho>$100 AU) containing a UCD companion are marked as five point stars and those highlighted in red also contain a VLM binary (hence at least a triple). Those marked as squares are systems containing a tight or widely separated UCD with an age $<$ 500 Myr.  Objects marked by open circles come from stellar companion catalogs.   The NLTT 20346/2MASSJ0850+1057 system is the enlarged five-point star. The minimum binding energy corresponding to tight very low mass systems is labeled (see \citealt{2007ApJ...660.1492C,2003ApJ...587..407C} and \citealt{2003ApJ...586..512B}). } 
\label{fig:be2}
\end{figure}

We have added NLTT 20346/2MASS J0850+1057 to the growing list of VLM (M$_{tot}<$0.2~M$_{\sun}$) multiples widely separated from a more massive companion. Table ~\ref{system} is a selected compilation of these widely-separated multiples and demonstrates a range in both mass and separation of components.  Within this list, NLTT 20346/2MASS J0850+1057 has a significantly lower binding energy and is one of the lowest total mass triple, quadruple, and/or quintuple systems known.  Figure ~\ref{fig:be2} shows the binding energy versus total mass of systems gathered from the literature\footnote{Stellar companions were gathered from the catalogs of \citealt{1991A&A...248..485D}, \citealt{1992ApJ...396..178F}, and \citealt{1997A&AS..124...75T}; and young UCD companion systems from  \citealt{2005ApJ...633..452K, 2006ApJ...649..306K}, \citealt{2007ApJ...663..394K}, \citealt{2009ApJ...691.1265L}, and \citet{2009ApJ...697..824A}. Details on the field UCD systems were gathered from the Very Low Mass Binary Archive ($http://vlmbinaries.org$; see \citealt{2007prpl.conf..427B} and references therein.)}   and demonstrates that this new system falls well below the binding energy limitation set by known tight low mass (M$_{tot}$ $<$ 0.2 M$_{\sun}$) multiples (see \citealt{2007ApJ...660.1492C,2003ApJ...587..407C} and \citealt{2003ApJ...586..512B}).   For comparison, \citet{1998MNRAS.299..955P} use the virial theorem to obtain the gravitational binding energy of the Pleiades cluster.  Their  derived value ($\sim$ 5.3 x 10$^{38}$ erg) is smaller than the estimated binding energy for the NLTT 20346/2MASS J0850+1057 system (and smaller than the lowest binding energy systems found to date), although it is unclear how long this association will remain bound.  An interesting investigation would be to search for stars in the vicinity of NLTT 20346/2MASS J0850+1057 as the hierarchical nature of this system combined with the common kinematics, potential youth, and wide separation between the M dwarf and brown dwarf systems might be indicative of its own moving group.  

\begin{deluxetable*}{llllllllllllllllllllllll}
\tabletypesize{\footnotesize}
\label{tab:tab1}

\tablecaption{Details on Triple/Quadruple Systems Containing at Least Two Ultracool Dwarfs \label{system}}
\tablewidth{0pt}
\tiny
\tablehead{
\colhead{Name} &
\colhead{SpT} &
\colhead{SpT} &
\colhead{$\rho$}  &
\colhead{$\rho$} &
\colhead{$\rho$} &
\colhead{$\rho$} &
\colhead{M}&
\colhead{M} &
\colhead{M }&
\colhead{q\tablenotemark{a}} &
\colhead{BE} &
\colhead{Ref}\\
 &
&
 &
\colhead{($\arcsec$)} &
\colhead{(AU)}&
\colhead{($\arcsec$)} &
\colhead{(AU)}&
\colhead{($M_{\sun}$)} &
\colhead{($M_{\sun}$)} &
\colhead{($M_{\sun}$)}&
&
\colhead{10$^{41}$ Erg}\\
 &
\colhead{\tiny{Primary}} &
\colhead{\tiny{Secondary}}&
\colhead{\tiny {star-UCD}} &
\colhead{\tiny {star-UCD}}&
\colhead{\tiny{UCD-UCD}}&
\colhead{\tiny{UCD-UCD}} &
\colhead{\tiny{Primary}}&
\colhead{\tiny{Secondary}}&
\colhead{\tiny{Total}}&
&
\\
\colhead{(1)}  &
\colhead{(2)}  &
\colhead{(3)}  &
\colhead{(4)}  &
\colhead{(5)}  &
\colhead{(6)}  &
\colhead{(7)}  &
\colhead{(8)}  &
\colhead{(9)}  &
\colhead{(10)}  &
\colhead{(11)}  &
\colhead{(12)}  &
\colhead{(13)}  \\}
\startdata
NLTT 20346			&	M5+M6	&	L7+L6.5		&	248	&	7700		&	0.16	&	4.7	&	0.290	&	0.070	&	0.360	&	0.24	&	0.37	& 1A,1B\\		\hline
G 171-58	&	F8		&	L4+L4		&	218	&	9202		&	0.33	&	10.2	&	1.150	&	0.095	&	1.245	&	0.08	&	1.7	& 2\\
HD 221356	&	F8		&	M8+L3		&	452	&	11900	&	0.57	&	15	&	1.020	&	0.160	&	1.180	&	0.16	&	1.9	& 3\\
G 124-62	&	dM4.5e	&	L1+L1		&	44	&	1496		&	0.42	&	14.3	&	0.210	&	0.144	&	0.354	&	0.69	&	2.8	& 4\\
eps Ind				&	K5		&	T1+T6		&	402	&	1460		&	0.73	&	2.6	&	0.670	&	0.072	&	0.742	&	0.11	&	4.6	& 5\\
Gl 417	&	G0+G0	&	L4.5+L6		&	90	&	2000		&	0.07	&	1.5	&	0.940	&	0.143	&	1.083	&	0.15	&	9.4	& 6A,6B\\
LP 213-67	&	M6.5		&	M8+L0		&	14	&	230		&	0.12	&	2.8	&	0.100	&	0.176	&	0.276	&	1.76	&	11	& 7A,7B\\
GJ 1001	&	M4		&	L4.5+L4.5		&	19	&	180		&	0.09	&	1	&	0.250	&	0.136	&	0.386	&	0.54	&	26	& 8A,8B,8C\\
Gl337					&	G8+K1	&	L8+L8/T		&	43	&	880		&	0.53	&	10.9	&	1.740	&	0.110	&	1.850	&	0.06	&	30	& 9A,9B\\
HD65216				&	G5		&	M7+L2		&	7	&	253		&	0.17	&	5.9	&	0.940	&	0.167	&	1.107	&	0.18	&	87	& 10\\
GJ569					&	M2.5		&	M9.0+M9.0	&	5	&	50		&	0.10	&	0.9	&	0.350	&	0.126	&	0.476	&	0.36	&	123	& 11\\
Kelu-1					&	L0.5/T7.5	&	L3p			&	--	&	--		&	0.29	&	5.4	&	0.125	&	0.055	&	0.180	&	0.44	&	178	& 12A,12B,12C\\
LHS1070				&	M5.5		&	M8.5+M9.0	&	1	&	4		&	0.45	&	3.4	&	0.210	&	0.138	&	0.348	&	0.66	&	1134	 & 13A,13B\\
HD130948				&	G2		&	L4+L4		&	0.4	&	7		&	0.13	&	2.4	&	1.030	&	0.109	&	1.139	&	0.11	&	2178	 & 14A,14B\\

\enddata
\tablerefs{1A=This Paper 1B=Burgasser et al (2010)  2=\citet{2010AJ....139..176F} 3=\citet{2007ApJ...667..520C} 4=\citet{2005A&A...440..967S} 5=\citet{2003A&A...398L..29S} 6A=\citet{2001AJ....121.3235K} 6B=\citet{2003AJ....126.1526B}
7A=\citet{2000MNRAS.311..385G} 7B=\citet{2003ApJ...587..407C} 8A=\citet{2004AJ....128.1733G} 8B=\citet{1999ApJ...519..802K} 8C=\citet{1999Sci...283.1718M} 9A=\citet{2001AJ....122.1989W} 9B=\citet{2005AJ....129.2849B} 10=\citet{2007MNRAS.378.1328M} 11=\citet{2000ApJ...529L..37M} 12A=\citet{1997ApJ...491L.107R} 12B=\citet{2005ApJ...634..616L} 12C=\citet{2008arXiv0811.0556S} 13A=\citet{1994A&A...291L..47L} 13B=\citet{2001A&A...367..183L} 14A=\cite{2002ApJ...567L.133P} 14B=\citet{2008arXiv0807.2450D}}
\tablenotetext{a}{q represents the mass ratio $\frac{M_{secondary}}{M_{primary}}$}

\end{deluxetable*}

\begin{figure}[htbp]
\begin{center}
\includegraphics[width=1.0\hsize]{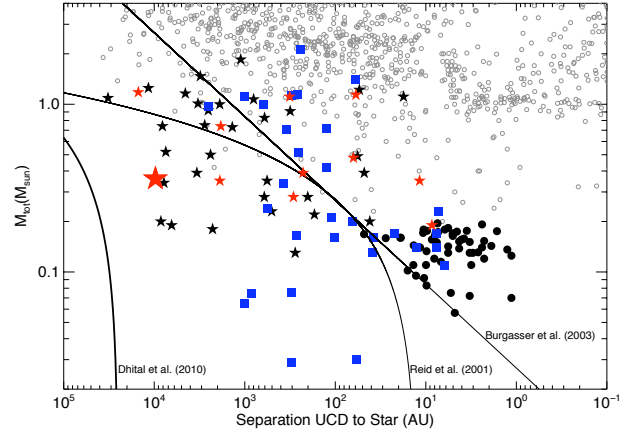}
\end{center}
\caption{A plot of the total mass vs. separation.  Symbols are described in Figure ~\ref{fig:be2}. We have over-plotted the \citet{2001AJ....121..489R} curve (center) which distinguished the cut-off for the formation of wide stellar pairs as well as the \citet{2003ApJ...586..512B} line (far right--which is specific for M$_{tot}<$0.2 M$_{\sun}$ field systems) and the log-normal limitation found in \citet{2010AJ....139.2566D} (far left--for systems with M$_{tot}$ $>$ 0.3 M$_{\sun}$).} 
\label{fig:be}
\end{figure}

Figure ~\ref{fig:be} shows the M$_{tot}$ vs. separation for the same companion systems. The studies of \citet{2001AJ....121..489R} and \citet{2003ApJ...586..512B} suggested an empirical limit for the stability of VLM multiples based on objects that were known at the time of the respective works.  Neither cut-off seems appropriate for NLTT 20346/2MASS J0850+1057 or the collection of slightly more massive widely separated systems plotted on Figure ~\ref{fig:be} (see discussions in \citealt{2010AJ....139..176F} and \citealt{2010AJ....139.2566D}).  \citet{2010AJ....139.2566D} found a log-normal limitation on the separation of binaries in their catalog of 1342 wide ($>$ 500 AU) low-mass (majority had M$_{tot}$  $>$ 0.3) systems and this cut-off  does encompass NLTT 20346/2MASS J0850AB and all low mass objects on Figure ~\ref{fig:be}.  

The stability of binary systems will be a function of age and total mass (see \citealt{1987ApJ...312..367W}).  While NLTT 20346/2MASS J0850AB is fit by the cut-off limitation of higher mass companions, it is significantly different from other known slightly less massive higher order multiples so we investigate one possible mechanism for its formation.  Recent work by \citet{2009MNRAS.392..413S} has been successful in accounting for widely separated VLM binaries using simulations of gravitational fragmentation of massive extended disks. In their smoothed particle hydrodynamic (SPH) simulations, a system with M$_{DISK}$=0.7M$_{\sun}$, R$_{DISK}$=400AU, M$_{star}$=0.7 M$_{\sun}$ is evolved for up to 20 kyr followed by an N-body dynamical evolution for up to 200 Kyr.  After 12 simulations, 96 stars are formed with brown dwarf or low--mass secondaries and among those companions,  9 are tight VLM multiples.  The characteristics of the triple systems were listed in  \citet{2009MNRAS.392..413S} and one was found to have a total secondary mass of 0.148 M$_{\sun}$, a close brown dwarf-brown dwarf binary separation of $\sim$ 1 AU and a wide separation from the central star of 7700 AU.  This is slightly more massive but highly analogous to our proposed system indicating that gravitational fragmentation could account for the existence of NLTT 20346/2MASS J0850+1057.  We found no analogs created using different formation mechanisms therefore comparisons with the predictions from ejection, etc. are not possible at this time.

The simulations of  \citet{2009arXiv0911.3662S} showed a significant population of low mass (M$_{secondary}$ $<$ 100 M$_{J}$) companions  at distances out to 10,000 AU.  This is in agreement with recent studies that have revealed a growing number of ultracool dwarfs with separations approaching and in some cases exceeding 10,000 AU from a companion (see Table 1 in \citealt{2009AJ....137....1F} and references there-in; \citealt{2010AJ....139.2566D};\citealt{2010MNRAS.404.1817Z}).   The caveat in this comparison is that the observed sample covers primarily field age objects (1-5 Gyr)  while the  \citet{2009arXiv0911.3662S}  simulations stop after only 200 kyr of dynamical evolution.  The existence of a population of older, widely separated systems suggests that  dynamical interactions into field ages does not disrupt all systems out to 10,000 AU.  


\citet{2009MNRAS.392..413S} find that the nine low-mass binaries formed through gravitational fragmentation that remain bound to their parent star have high eccentricities ($<$e$_{BIN}$$>$ of  0.7 $\pm$ 0.2).  They postulate that the dynamical interactions which form them, and/or subsequent dynamical interactions with other stars that have condensed out of the disk, account for the increased eccentricity of the orbit.  The masses  of the primary stars in that work are not reported although most are described as Sun-like.  We can compare the eccentricities of known widely separated VLM binaries with the simulation predictions although the primary masses are most likely smaller.  Three of the VLM binary systems from Table ~\ref{system}, GJ 569B, HD 130948B, and 2MASSJ 0850+1057, have dynamical mass measurements and eccentricities measured (\citealt{2008arXiv0807.2450D}; \citealt{2010arXiv1001.4800K}).  The mean eccentricity for these three systems is 0.4 $\pm$ 0.2, falling within 1$\sigma$ of the simulations.  Further information on the eccentricity of VLM binaries widely separated from a more massive component will enhance our understanding of dynamical interactions of low-mass hierarchical systems.

\section{SUMMARY}
We have presented evidence that NLTT 20346 is an M5+M6 binary, and it forms a comoving wide ( $\sim$7700 AU projected separation) companion system with 2MASS J0850+1057 at a distance of $\sim$ 31 pc from the Sun.  NLTT 20346 has moderate chromospheric emission and a statistical age based on H$_{\alpha}$ activity of $\sim$5 - 7 Gyr for both components.  However, it is also an X-ray source and has an age of 250-450 Myr based on X-ray activity. 2MASS J0850+1057 shows strong Li absorption indicating an upper age limit between 0.8-1.5 Gyr, favoring the younger age for the primary.   Assuming an age of 0.25-1.5 Gyr for the quadruple, we calculate a total mass for 2MASS J0850+1057 of 0.07 $\pm$0.02M$_{\sun}$ and a total mass for NLTT 20346 of 0.29$\pm$0.01 M$_{\sun}$.  The new age will be an important parameter for refining dynamical mass calculations of 2MASS J0850+1057 (\citealt{2010arXiv1001.4800K}) so this object can be used as further evidence for or against the conclusion that evolutionary models are under-estimating the masses for L dwarfs. 

We report a new parallax measurement for 2MASS J0850+1057 based on mulit-epoch ISPI imaging.  Two prior measurements from  \citet{2002AJ....124.1170D} and \citet{2004AJ....127.2948V} were discrepant from one another showing a 2.4$\sigma$ difference in parallactic angle and a 3.4$\sigma$ difference in proper motion position angle. Our new measurement is too uncertain to differentiate between the two previous works but a contaminating object located on our recent astrometric follow-up images may provide an explanation for the differing distance measurements. 

The binding energy for this new quadruple is among the lowest known for a wide companion system.  \citet{2009MNRAS.392..413S} presented an analogous system to NLTT 20346/2MASS J0850+1057 formed by the gravitational fragmentation of a massive extended disk.  In that work, they create a significant population of wide ($\rho$ $>>$ 100 AU) systems containing low mass (M$_{secondary}$ $<$ 50 M$_{J}$) secondaries analogous to wide companion systems now being found.   While dynamical evolution of their systems ended after just 200 kyr, the majority of the wide systems found to date are field aged indicating that such systems remain stable for several Gyr.   


\acknowledgments{We thank the anonymous referee for extremely useful comments and suggestions as well as E. Shkolnik for providing activity data on a sample of field and cluster M dwarfs, and M. Shara and F. Walter for useful comments on an early draft.  This publication has made use of the VLM Binaries Archive maintained by Nick Siegler at http://www.vlmbinaries.org as well as the data products from the Two Micron All-Sky Survey, which is a joint project of the University of Massachusetts and the Infrared Processing and Analysis Center/California Institute of Technology, funded by the National Aeronautics and Space Administration and the National Science Foundation. This research has made use of the NASA/ IPAC Infrared Science Archive, which is operated by the Jet Propulsion Laboratory, California Institute of Technology, under contract with the National Aeronautics and Space Administration. We acknowledge receipt of observation time through NOAO.  We also thank 4.0m telescope operators C. Aguilera, M. Gonzalez, and A. Alvarez. J. Faherty gratefully acknowledges support from Hilary Lipsitz and from the AMNH. }

\clearpage
\bibliographystyle{apj}
\bibliography{paper}

\end{document}